# Stochastic electron heating in the laser and quasi-static electric and magnetic fields

Yanzeng Zhang, S. I. Krasheninnikov, Alexey Knyazev
University of California San Diego, 9500 Gilman Dr., La Jolla, CA 92095

E-mail: yaz148@eng.ucsd.edu

**Abstract.**

The dynamics of relativistic electrons in the intense laser radiation and quasi-static electromagnetic fields both along and across to the laser propagating direction are studied in the 3/2 dimensional Hamiltonian framework. It is shown that the unperturbed oscillations of the relativistic electron in these electric fields could exhibit a long tail of harmonics which makes an onset of stochastic electron motion be a primary candidate for electron heating. The Poincaré mappings describing the electron motions in the laser and electric fields only are derived from which the criterions for instability are obtained. It follows that for both transverse and longitudinal electric fields, there exist upper limits of the stochastic electron energy depending on the laser intensity and electric field strength. Specifically, these maximum stochastic energies are enhanced by a strong laser intensity but weak electric field. Such stochastic heating would be reduced by the superluminal phase velocity in both cases. The impacts of the magnetic fields on the electron dynamics are different for these two cases and discussed qualitatively. These analytic results are confirmed by the numerical simulations of solving the 3/2D Hamiltonian equations directly.

## I. Introduction

The ability of the interaction of intense laser radiation with plasma to generate highly energetic electrons is one of the most interesting features in laser-plasma physics that has great potentials for many applications (including ion acceleration, X-ray generation, positron production, etc.). The mechanisms of electron heating to obtain high energy have been suggested and studied analytically, numerically, and experimentally (see e.g. Refs. 1-33 and the references therein) over many years. Many of these works (see e.g. Refs. 3, 7-10, 14, 20-33) reveal that the presence of self-generated or externally applied quasi-static electric and magnetic fields could significantly increase the electron energy gained from the laser well beyond the ponderomotive scaling[6].

However, such synergistic effects can be due to different mechanisms. In Ref. 7 the synergy was attributed to the betatron resonance where the energetic electron undergoes betatron oscillation in the static fields along the polarization of the laser pulse electric field. When the betatron frequency is close to the laser frequency as witnessed by the relativistic electron, an efficient energy exchange is possible. Refs. 23-24 showed that the strong laser field could nonlinearly modulate the betatron oscillation frequency to enable the parametric amplification regardless of the laser polarization direction. In Ref. 21 it was shown that the synergistic effects presented in the longitudinal electric field is related[26,27] to an onset of stochasticity of the electron motion (see e.g. Refs. 34-36) which is reduced to the Poincaré mapping for the case of a V-shape electrostatic potential, $U(z) \propto |z|$, where z directed the laser propagation while the stochastic acceleration of the particle in the field of electromagnetic wave and a constant transverse magnetic field was studied in Ref. 37. The impact of a longitudinal decelerating electric field on electron acceleration in the laser and transverse electric field was investigated in Ref. 38 where the decelerating electric field help with maintaining high-amplitude betatron oscillations.

Although starting from different considerations, the maximum electron energy has been estimated as a function of the parameters of the laser radiation and static electric and magnetic fields[23,24,26,28,29] through simplified analyses and particle-in-cell (PIC) simulations, the identification of the synergistic mechanism(-s) is poorly resolved due to the strong nonlinearity of the relativistic electron motion in these fields.

Recently, it was shown that the electron dynamics can be described by the 3/2 dimensional (3/2D) Hamiltonian approach for a homogeneous magnetic field and linearly polarized laser plane wave[39]. Refs. 40, 41 extended this approach to the cases of arbitrarily polarized laser radiation depending only on the phase variable $(v_{ph}t - z)/\lambda$ and arbitrary quasi-static electric and magnetic fields in the directions both along and across to the laser propagation direction. Here $\lambda$ is the laser pulse wavelength; $v_{ph} \geq c$ is the phase velocity; t and z are, respectively, the time and coordinate along the laser propagating direction. This method has greatly simplified the analysis of electron dynamics benefitting from the fundamental results of previous studies on regular and stochastic motion in Hamiltonian systems (see e.g. Refs. 34-36 and the references therein). The numerical simulations[40,41] demonstrated that in the presence of quasi-static electromagnetic fields, an onset of stochasticity is accounted for the electron heating which is depressed by the superluminal phase velocity. However, the mechanism of the stochastic heating is still not clear.

In the present paper, the dynamics of the relativistic electrons in the laser and quasi-static electromagnetic fields is studied in detail within the 3/2D Hamiltonian framework. For both the cases of transverse and longitudinal electric fields, the 3/2D Hamiltonians are the total electron energy in the static fields, which will be conserved when electron oscillates in the static fields without laser radiation (hereafter the electron motion without laser is referred as unperturbed problem). Like the nonlinear pendulum problem[35], the electron coordinates and momenta in such oscillating motions have harmonics with a long tail distribution for relativistic electrons. As a result the resonances of the electron motion frequency with the laser frequency could be largely broadened and thus overlap which causes an onset of stochastic electron heating.

The role of the static fields in this mechanism is to reduce the longitudinal dephasing rate $\gamma - p_z / m_e c$ between the electron and laser beam such that the electron could stay in phase with the laser and effectively exchange energy with laser, instead of directly transferring substantial energy to the electron[11]. Here $\gamma$ is the relativistic factor and $p_z$ and $m_e$ are the electron momentum along laser propagation direction and mass, respectively. The smallest dephasing rate $\gamma - p_z / m_e c$ is denoting the strongest interaction between the electron and laser wave which is also called the nonadiabatic interaction or "collisions" in the rest of paper. Following Ref. 37 we would evaluate the change of the electron energy between two consecutive collisions by using the stationary-phase method where the unperturbed electron trajectory is applied under the assumption that the energy change during each collision is small compared with the electron energy itself.

We introduce the normalized variables $\hat{t} = t\omega$, $\hat{\vec{r}} = k\vec{r}$, $\hat{\vec{v}} = \vec{v}/c$, $(\hat{\tilde{\vec{A}}}, \hat{\vec{A}}_B) = e(\tilde{\vec{A}}, \vec{A}_B)/m_e c^2$ and $\hat{U} = eU/m_e c^3$, where $\omega$ and $k$ are, respectively, the frequency and wave number of the laser radiation described by the vector potential $\tilde{\vec{A}}$; $U$ and $A_B$ are the quasi-static electric and magnetic vector potentials, respectively. In the rest of this letter, the "hats" over the dimensionless quantities will be omitted to simplify our expressions. For the sake of simplicity, we will use plane wave of laser radiation which has been justified by the numerical simulations (e.g. Ref. 41) where the laser field in the ion channel has a planar structure with superluminal phase velocity. As a result, the laser can be described by $\tilde{\vec{A}} = \tilde{A}_x(t - z/\alpha)\vec{e}_x + \tilde{A}_y(t - z/\alpha)\vec{e}_y$, where $\tilde{A}_x$ and $\tilde{A}_y$ are used to distinguish the laser polarizations and $\alpha = v_{ph}/c \geq 1$ is the normalized laser phase velocity.

For both cases, we would start with general 3/2D Hamiltonians for arbitrarily polarized laser with superluminal phase velocity and quasi-static electric and magnetic fields but the impacts of the electric and magnetic field are discussed separately. The rest of the paper is organized as follows. In Sec. II we examine the electron dynamics in the transverse electric field and laser radiation where the quadratic form of the electric potential would be used[7]. The role of the longitudinal electric field is examined in Sec. III for general power form of $U(z) \propto k_u |z|^p$, where p=1 was studied in Refs. 26, 27. However, we will show that the electron dynamics for $p > 1$ is quite different from that for p=1. Sec. IV will discuss the impacts of the magnetic fields on these electron dynamics. Numerical simulations are presented in Sec. V while in Sec. VI we will summarize and discuss the main results of this paper.

**II. Stochastic electric motion in the laser and transverse electric fields**

We examine electron dynamics in the transverse electromagnetic fields described by the potentials of $U = \kappa_u y^2/2$ and $A_B = \kappa_b y^2/2$ where $\kappa_u$ and $\kappa_b$ are constants denoting, respectively, the electric and magnetic fields strength. Note that the electric field is along y-direction while the magnetic field is in the x-direction. This system contains two integrals of motion as $\bar{P}_x \equiv p_x - \tilde{A}_x$ and $C_\perp \equiv \gamma - \alpha p_z + W_\alpha^B$ where $W_\alpha^B = U(y) + \alpha A_B(y)$. By choosing proper variables $\xi = t - z/\alpha$ and $\tilde{p}_y = p_y - \tilde{A}_y$, the electron satisfies the following 3/2D Hamiltonian equations[41]

$$\frac{d\tilde{p}_y}{d\xi} = -\frac{\partial H_y^\alpha}{\partial y}, \text{ and } \frac{dy}{d\xi} = \frac{\partial H_y^\alpha}{\partial \tilde{p}_y}, \tag{1}$$

where

$$H_y^\alpha(\tilde{p}_y, y, \xi) = \frac{\alpha}{\alpha^2 - 1}\left\{\sqrt{(C_\perp - W_\alpha^B)^2 + (\alpha^2 - 1)P_{\perp,y}^2} + W_\alpha^U - \frac{C_\perp}{\alpha}\right\} = \gamma + U = E, \tag{2}$$

$W_\alpha^U = \alpha U(y) + A_B(y)$ and $P_{\perp,y}^2 = 1 + (\bar{P}_x + \tilde{A}_x)^2 + (\tilde{p}_y + \tilde{A}_y)^2$.

The nonadiabatic region corresponds to the minimum of the dephasing rate $\gamma - p_z = C_\perp - W_\alpha^B(y) + (\alpha - 1)p_z$, which indicates that the presence of the static fields could enhance the electron-laser interaction when $W_\alpha^B(y)$ approaches $C_\perp$ if $\kappa_u + \alpha\kappa_b > 0$ while the superluminal phase velocity of laser radiation, $\alpha > 1$, would reduce it. For simplicity, in what follows, we will consider the luminal case ($\alpha = 1$) where Eq. (2) reverts to

$$H_y = \frac{1}{2}\left\{\frac{1 + (\bar{P}_x + \tilde{A}_x)^2 + (\tilde{p}_y + \tilde{A}_y)^2}{C_\perp - W^{(+)}(y)} + W^{(-)}(y) + C_\perp\right\} = \gamma + U = E, \tag{3}$$

where $W^{(\pm)} = U(y) \pm A_B(y)$. Then in order to have strong electron laser interactions, we would consider $\kappa_u + \kappa_b > 0$ and Eq. (3) indicates that such strong interaction occurs at small denominator of $C_\perp - W^{(+)}(y) = \gamma - p_z$. For relativistic electrons with energy $E \gg W^{(-)}(y_{max})$, where $y_{max} \approx 2C_\perp/(\kappa_u + \kappa_b)$, the impact of the $W^{(-)}(y)$ is negligible in the nonadiabatic region. As a result, the electric and magnetic fields play similar role in the electron dynamics and we can ignore the magnetic field, taking $\kappa_u + \kappa_b$ as the "effective" electric coefficient. In the rest of this section we would take $\kappa_u > 0$ and $\kappa_b = 0$.

First we consider electron trajectory without the perturbation of laser radiation, where the electron energy is conserved. From Eq. (3), we see that the electron motion is bounded between $y_{max,min} \approx \pm\sqrt{2C_\perp \kappa_u^{-1}}$ and we find

$$\tilde{p}_y = -\sqrt{2EC_\perp}\sin\theta, \ y = \sqrt{2C_\perp \kappa_u^{-1}}\cos\theta, \ d\theta/d\xi = \sqrt{E\kappa_u}C_\perp^{-1}\sin^{-2}\theta, \tag{4}$$

where the angle $\theta$ goes clockwise direction (seen Fig. 1) with $\theta = 0$ ($\theta = \pi$) corresponding to $y_{max}$ ($y_{min}$). The last expression in Eq. (4) reads

$$\zeta \equiv \xi - \xi_n = \frac{C_\perp(2\theta - \sin 2\theta)}{4\sqrt{E\kappa_u}}, \tag{5}$$

where $\xi_{2n}$ is the time of the *n*th passage of the electron through $y_{max}$. Substituting $\theta = 2\pi$ in the above expression provides the unperturbed electron frequency over closed orbit

$$\Omega = \frac{2\sqrt{E\kappa_u}}{C_\perp}, \tag{6}$$

The same result could be derived from the calculation of the action, I, of the electron from Eq. (4), which gives $I = \oint \tilde{p}_y dy / 2\pi = C_\perp \kappa_u^{-1/2} E^{1/2}$ and $\Omega = \partial E / \partial I$ results in the expression (6).

However, from Eq. (3) we find that an impact of laser radiation on electron trajectory could be largely ignored (with exception rather narrow regions in the vicinity of $\theta = 0$ and $\theta = \pi$, where $\tilde{p}_y$ approaches zero) only for the energies

$$E > E_{pond} = a_0^2 / 2C_\perp, \tag{7}$$

where $E_{pond}$ in our case could be considered as a ponderomotive energy scaling. For such electron energy, the unperturbed electron orbit in Eq. (4), even though approximates to an ellipse, is significantly stretched along $\tilde{p}_y$. As a result, its dynamics is to some extent similar to that of oscillating particles in a "square well"[35]. It follows that the zigzag-like dependence of y on the time $\xi$ and "step-shape" of $\tilde{p}_y$ like the particle trajectory in the square well would have harmonics with long tail distribution (the amplitude of the *n*th harmonic is approximately to $a_n \sim 1/n$). As a result, for efficient electron-laser interactions the resonances of electron frequency $\Omega$ (or its harmonics) with laser frequency (which is unity in our normalization and thus $n\Omega = 1$ where $n \geq 1$ is the integer number) could be largely broadened and overlap, which causes stochastic electron heating [35]. Thus, the requirement $\Omega \leq 1$ sets the limit for efficient electron-laser interactions since at $\Omega > 1$ an impact of laser field becomes weak and electrons do regular motion. As a result, an absolute maximum energy where electron can be heated up via interactions with the laser radiation and the transverse electric field is considered as

$$E_{max}^{abs} = C_\perp^2 / 4\kappa_u. \tag{8}$$

From Eqs. (7, 8) we find that in order to heat electron beyond ponderomotive scaling, we should have

$$\varepsilon \equiv \sqrt{2} a_0 \kappa_u^{1/2} C_\perp^{-3/2} \ll 1, \tag{9}$$

which appears in Ref. 23 for the parametric amplification of laser-drive electron acceleration and Ref. 29 as result of time-average theory (the former considered large $\varepsilon$ where $\varepsilon > \varepsilon_* \approx 10.2$ will cause instability and accelerate electrons, while the latter focused on small $\varepsilon$). Then we have $E_{max}^{abs} = E_{pond} \varepsilon^{-2}$ and thus, in what follows, we would use $E_{max}^{abs}$ to scale the electron energies.

Considering the Hamiltonian property, the energy change of the electron for the polarization of $\vec{\tilde{A}} = a_0 \sin(\xi)\vec{e}_y$ and $\vec{\tilde{A}} = a_0 \sin(\xi)\vec{e}_x$ would be given, respectively, by

$$E_y(\xi) - E_y(\xi_{max}) = \int_{\xi_{max}}^{\xi} \frac{a_0(a_0 \sin\xi + \tilde{p}_y)\cos\xi}{C_\perp - U(y)} d\xi, \tag{10a}$$

and

$$E_x(\xi) - E_x(\xi_{max}) = \int_{\xi_{max}}^{\xi} \frac{a_0^2 \sin\xi \cos\xi}{C_\perp - U(y)} d\xi. \tag{10b}$$

We consider the case $\Omega \ll 1$ and thus electron heating is due to overlapping of high-n resonances ($n\Omega = 1$). Then the electron energy changes during relatively short time of strong interactions of electron with laser radiation in the vicinity of $y = y_{max}$ and $y = y_{min}$. We assume that electron energy variation during such collisions, $\Delta E$, is small $|\Delta E| \ll E$ such that the unperturbed electron trajectories in equations (4) can be applied to assess the electron energy change between two consecutive collisions from Eq. (10), i.e.,

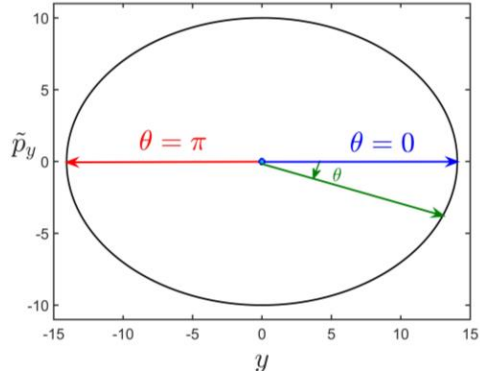

FIG. 1. Schematic view of the ultrarelativistic electron trajectory in the transverse electric field for $C_\perp = 1$, $\kappa_u = 0.01$ and $E = 50$.

$$\Delta E_y = a_0 \sqrt{2C_\perp / \kappa_u} \sin\xi_j \int_{-\pi/2}^{\pi/2} \sin\theta \sin\left[\Omega^{-1}\left(\theta - \frac{\sin 2\theta}{2}\right)\right] d\theta$$
$$+ \frac{a_0^2}{2\sqrt{E\kappa_u}} \sin(2\xi_j) \int_{-\pi/2}^{\pi/2} \cos\left[2\Omega^{-1}\left(\theta - \frac{\sin 2\theta}{2}\right)\right] d\theta \quad (11a)$$

$$\Delta E_x = \frac{a_0^2}{2\sqrt{E\kappa_u}} \sin(2\xi_j) \int_{-\pi/2}^{\pi/2} \cos\left[2\Omega^{-1}\left(\theta - \frac{\sin 2\theta}{2}\right)\right] d\theta, \quad (11b)$$

where $\xi_j$ is the time of previous collision in the vicinity of $y_{min}$.

Under the condition of $\Omega \ll 1$, we notice the fact that $|\theta| \ll 1$ mostly contributes to the integrals in equation (11), which enables the Taylor expansion of the terms in the brackets of equations (11) as $\theta - \sin(2\theta)/2 = 2\theta^3/3$ and the integral limit can be extended to infinity. Then the integrals in Eq. (11) are degenerated to the Airy function Ai(x) and its first derivative Ai'(x) at zero, and after some algebra we obtain

$$\Delta E_x = 2^{1/3} 3^{-1/6} \Gamma(1/3) \left(E_{max}^{abs}\right)^{4/3} \varepsilon^2 E^{-1/3} \sin(2\xi_{2n}), \quad (12a)$$

$$\Delta E_y = 2^{4/3} 3^{1/6} \Gamma(2/3) \left(E_{max}^{abs}\right)^{2/3} \varepsilon E^{1/3} \sin(\xi_j) + \Delta E_x. \quad (12b)$$

where $\Gamma(x)$ is the gamma function. Then condition of $|\Delta E| \ll E$ requires that

$$E \gg E_{max}^{abs} \varepsilon^{3/2} = E_{pond} \varepsilon^{-1/2}. \quad (13)$$

Therefore, our analysis is only valid for the electron with energy of $\varepsilon^{3/2} \ll E/E_{max}^{abs} \ll 1$ under the condition of $\varepsilon \ll 1$. For electron with energy smaller than that in equation (13), the change of the electron orbit due to the electron laser interaction is large and our estimate in Eqs. (12) using the unperturbed electron orbit is not accurate. Noticing that these electrons could still undergo stochasticity and we are more interested in the maximum electron energy gain, we would consider the electron satisfying Eq. (13). One important result drawn from condition (13) is that the first term on the right hand side of Eq. (12b) is dominated over the second one such that

$$\Delta E_y \approx 2^{4/3} 3^{1/6} \Gamma(2/3) \left(E_{max}^{abs}\right)^{2/3} \varepsilon E^{1/3} \sin(\xi_j) \gg \Delta E_x . \tag{14}$$

By using the same procedure or considering symmetry of this system, we can show that the energy variation in the vicinity of $y_{min}$ is also given by expression (14) except a "-" sign in front for $\Delta E_y$ as the work done by the laser depends on the direction of $\tilde{p}_y$. Therefore, we can ignore the difference between $y_{max}$ and $y_{min}$ and obtain the Poincaré mappings from Eqs. (6, 12, 14) as

$$\Pi_{n+1}^{x,y} = \Pi_n^{x,y} + Q_{x,y} \sin \psi_n^{x,y}, \quad \psi_{n+1}^{x,y} = \psi_n^{x,y} + \Pi_{n+1}^{\alpha_{x,y}}, \tag{15}$$

where $\Pi_n^x = (2\pi)^{-8/3} (E_n / E_{max}^{abs})^{4/3}$, $\psi_n^x = 2\xi_n$, $Q_x = 2^{-1/3} 3^{-7/6} \Gamma(1/3) \pi^{-8/3} \varepsilon^2$, $\alpha_x = -3/8$ and $\Pi_n^y = (\pi)^{-4/3} (E_n / E_{max}^{abs})^{2/3}$, $\psi_n^y = \xi_n$, $Q_y = (-1)^n 2^{7/3} 3^{-5/6} \Gamma(2/3) \pi^{-4/3} \varepsilon$, $\alpha_y = -3/4$ are the quantities corresponding to different polarizations. It can be easily shown that the mappings in Eq. (15) are symplectic and thus conserve the phase volume.

To see the stochasticity boundary, the relation $K_{x,y} = \left| d\psi_{n+1}^{x,y} / d\psi_n^{x,y} - 1 \right| \geq 1$ should be satisfied[35-37], which yields

$$K_x = k_x \left(E_{max}^{abs} E^{-1}\right)^{11/6} \varepsilon^2 > 1 , \text{ and } K_y = k_y \left(E_{max}^{abs} E^{-1}\right)^{7/6} \varepsilon > 1, \tag{16}$$

where $k_x = 2^{1/3} 3^{-1/6} \pi \Gamma(1/3)$ and $k_y = 2^{1/3} 3^{1/6} \pi \Gamma(2/3)$ are the numerical factors order of unity. The satisfaction of relation in (16) leads to the mixing in phase space and gives the stochasticity criterion as a function of the electron energy E, laser field amplitude $a_0$ and the electric field strength $\kappa_u$ where we have disregarded the region of phase $\psi$ in which chaos appears. It follows that for both polarizations there exits upper limit of the stochastic heating energy as

$$E_{max}^x \sim E_{max}^{abs} \varepsilon^{12/11}, \text{ and } E_{max}^y \sim E_{max}^{abs} \varepsilon^{6/7} . \tag{17}$$

We see that the maximum stochastic energy in Eq. (17) are smaller than $E_{max}^{abs}$ but they are above the ponderomotive scaling under the condition of $\varepsilon \ll 1$. Also it shows that $E_{max}^y \gg E_{max}^x$ and thus the electrons can gain more energy in the case where the laser electric field along the quasi-static electric field than that across to it. This is not surprising because the electron transverse velocity antiparallel to the laser electric field is larger in the former case and thus if we choose large $\bar{P}_x$ in the latter this difference can be eliminated. Moreover, Eq. (17) indicates that the upper energy boundary is relaxed for weak electric field. These results are confirmed by the numerical simulations in Sec. V.

### III. Stochastic electric motion in the longitudinal electric field

In this section, we will study the stochastic electric motion in the longitudinal electric field described by the general power form, i.e., $U(z) = k_u |z|^p$. According to Ref. 41, the electron dynamics in such system with magnetic field depending on z and directing in x can be described by the following Hamiltonian equations

$$\frac{d\tilde{z}}{d\xi} = -\frac{\partial H_z}{\partial \delta}, \text{ and } \frac{d\delta}{d\xi} = \frac{\partial H_z}{\partial \tilde{z}} \tag{18}$$

where

$$H_z(\tilde{z},\delta,\xi) = \frac{1}{2}\left(\frac{P_{\perp,z}^2}{\delta} + \delta\right) + U(\tilde{z}) \equiv \gamma + U, \tag{19}$$

$\delta = \gamma - p_z$, $\xi = t - z$, $\tilde{z} = z$, and $P_{\perp,z}^2 = 1 + \left(\bar{P}_x + \tilde{A}_x(\xi,\tilde{z})\right)^2 + \left(\bar{P}_y + \tilde{A}_y(\xi,\tilde{z}) + A_B(\tilde{z})\right)^2$. The two integrals of motion are $\bar{P}_x = p_x - \tilde{A}_x$ and $\bar{P}_y \equiv p_y - \tilde{A}_y - A_B$ where laser wave is in the form of $\tilde{A}_{(...)} = \tilde{A}_{(...)}(\xi + (1 - 1/\alpha)\tilde{z})$. The effect of the magnetic field will be discussed in Sec. IV and we consider $A_B = 0$ in this section. As a result, the laser polarization in Eq. (19) is not important so that the use of $\tilde{\vec{A}} = a_0 \sin(\xi - (1 - 1/\alpha)\tilde{z})\vec{e}_x$ will be applied in the following and for the sake of simplicity, we shall assume $\bar{P}_x = \bar{P}_y = 0$.

From Eq. (19) we see that the strongest electron interaction with the laser locates at small $\delta = \gamma - p_z$ when the electron passes through the bottom of the electrostatic potential well $\tilde{z} = 0$. Therefore, whereas the transverse electric field reduces the electron dephasing rate directly when it departures from the bottom of the electrostatic potential well, the longitudinal electric field seems to only confine the electrons such that they could enter the nonadiabatic region multiple times.

By using the same method as in Sec. II, the energy exchange between two consecutive collisions between the electron and the laser wave is given by

$$\Delta E = \int_{\xi_n}^{\xi_{n+1}} \frac{dH_z}{d\xi} d\xi = \int_{\xi_n}^{\xi_{n+1}} \frac{a_0^2 \sin[2\xi + (1 - 1/\alpha)\tilde{z}]}{2\delta} d\xi, \tag{20}$$

where $\xi = \xi_n$ is the time of $n$th passage of the electron through the nonadiabatic region $\delta \approx \delta_{min}$. From Eq. (19) we see that $\tilde{z}$ can be large when electrons oscillate in the electric field. As a result the superluminal phase velocity $\alpha > 1$ could induce a fast oscillation to the integral in Eq. (20) and thus reduce the electron laser interactions. In the following we will consider the luminal case ($\alpha = 1$).

In the action-angle variables $(I, \vartheta)$ of the unperturbed particle, from Eq. (19) we obtain

$$I = \frac{2p}{\pi(p+1)} E^{1+1/p} k_u^{-1/p}. \tag{21}$$

Therefore, the electron oscillating frequency $\Omega = \partial E / \partial I$ in the electric field reads

$$\Omega = 2^{-1} \pi k_u^{1/p} E^{-1/p}. \tag{22}$$

As a result, for efficient electron-laser interactions we should have the resonances of electron frequency with the laser frequency, i.e., $n\Omega = 1$ where $n \geq 1$ is the integer number. Therefore, we would consider $\Omega \leq 1$ and thus $E \geq k_u$.

Recalling that for $p = 1$, from Eq. (18) we have $\delta = \delta_{min} + k_u(\xi - \xi_n)$ ($\delta_{min}$ denotes the minimum of the coordinate $\delta$ and thus the center of the nonadiabatic region), the integral in Eq. (20) could be easily estimated as shown in Refs. 26 and 27. Therefore, we would focus on the general case of $p > 1$. Analyzing Eq. (18) we know that of $d\tilde{z}/d\xi \gg 1$ and $d\delta/d\xi \ll 1$ in the vicinity of $(\delta = \delta_{min}, \tilde{z} = 0)$. As a result, the electron will leave the vicinity of $(\delta_{min}, \tilde{z} = 0)$ and reach $(\delta_{min}, \tilde{z}_{max})$ in an extreme short time period where $\tilde{z}_{max} \sim pE^{1/p}k_u^{-1/p}$ and the electron trajectory in the nonadiabatic region $\delta \lesssim 1$ can be largely described by $\delta \sim \delta_{min} + pk_u\tilde{z}_{max}^{p-1}(\xi - \xi_n)$ and $\tilde{z} \sim \tilde{z}_{max}$ (notice that $\delta \sim \delta_{min} + pk_u\tilde{z}_{max}^{p-1}(\xi - \xi_n)$ still holds for p=1 but not $\tilde{z} \sim \tilde{z}_{max}$). This argument has been confirmed by the numerical simulations (e.g., see Fig. 6 (b) in Sec. V). This situation is not surprising since the electron has relativistic speed around $\delta_{min}$ and thus moves fast toward $\tilde{z}_{max}$ where it shall get efficient deceleration. However, the impact of the laser radiation on such unperturbed electron trajectory can be neglected only for $E \gg a_0^2/2 \approx E_{pond}$. As a result, we consider only the electron with energy of $E \geq \max\{k_u, E_{pond}\}$.

The condition of $\delta \sim 1$ corresponds to $\xi - \xi_n \sim p^{-1}k_u^{-1/p}E^{1/p-1} \equiv \zeta_f$ can be seen as the boundary of the nonadiabatic region. As we can see, for $p > 1$, $\zeta_f \ll 1$ and thus the variation of the numerator in Eq. (20) can be negligible. As a result, we find

$$\Delta E = a_0^2 \sin 2\xi_n \int_0^{\zeta_f} \frac{1}{pk_u^{1/p}E^{1-1/p}\zeta + \delta_{min}} d\zeta = \frac{a_0^2 \Lambda \sin 2\xi_n}{pk_u^{1/p}E^{1-1/p}}, \tag{23}$$

where $\Lambda \sim \ln(E)$. We see that Eq. (23) is consistent with the result for $p = 1$ with a slight difference of the coefficient[26, 27].

Eqs. (22) and (23) form the following Poincaré mapping

$$\Pi_{n+1} = \Pi_n + Q\sin\psi_n, \quad \psi_{n+1} = \psi_n + \Pi_{n+1}^{1/(2p-1)}, \tag{24}$$

where $\Pi_n = 8^{2p-1}k_u^{1/p-2}E_n^{2-1/p}$, $\psi_n = 2\xi_n$ and $Q = (2p-1)8^{2p-1}k_u^{-2}\Lambda a_0^2/p^2$. An onset of stochasticity thus requires

$$K = 8p^{-2}k_u^{-2/p}\Lambda a_0^2 E^{2(1-p)/p} \geq 1. \tag{25}$$

We can see that for $p > 1$, an upper boundary of the stochastic energy exits as

$$E_{max}^L \propto k_u^{-1/(p-1)}a_0^{p/(p-1)}, \tag{26}$$

which is quite different from the $p = 1$ case where there is no energy limit under the stochastic condition[26, 27] $a_0 > k_u$. Like the transverse case, higher energy boundary can be achieved at smaller $k_u$. The resonant condition of $\Omega < 1$ ($E > k_u$) now requires $a_0 > k_u$ while for electron stochastic heating above the ponderomotive scaling requires $a_0^{p-2}k_u < 1$. Therefore, we are interested in the region of $k_u < \min\{a_0, a_0^{2-p}\}$. The numerical results will be presented in Sec. V.

### IV. The effects of the magnetic fields

In the previous sections, the laser induced stochastic electron accelerations in the presence of quasi-static electric fields have been studied from the point of view of reduced Poincaré mapping. The impact of the magnetic field included in Eqs. (3, 19) for the luminal case still remains unsolved and will be discussed in this section. Notice that the role of the constant transverse magnetic field has been examined in Ref. 37 where the Poincaré mapping indicates that no upper limit of the stochastic electron heating exits for the luminal case. This Poincaré mapping can be easily recovered in the present 3/2D Hamiltonian language.

For the transverse electric field case, $\kappa_u + \kappa_b$ acts as the "effective" electric strength and we can get rid of the magnetic field for the same dependence of U and $A_B$ on the coordinate y. This has been verified by the numerical simulations. As a result, we could conclude that: if $\kappa_b > 0$, the magnetic field would weaken the stochastic electron motion by increasing the "effective" electric strength; if $\kappa_b < 0$ but $|\kappa_b| < \kappa_u$, it will enhance it; while the stochasticity no longer exists when $\kappa_u + \kappa_b < 0$.

However, the magnetic field plays quite different role in the electron motion for the longitudinal case and the laser polarization becomes important as shown in the following. For the laser polarized along the magnetic field (x-direction), the stochastic electron motion would always be suppressed. By analyzing Eq. (19) with magnetic field, we find that the electron will have a different unperturbed orbit $\delta_B(\tilde{z}) > \delta(\tilde{z})$ in the nonadiabatic region, where $\delta_B(\tilde{z})$ ($\delta(\tilde{z})$) is the quantity with (without) magnetic field for the same energy E. Then the energy variation in Eq. (20) for the laser wave polarized in x-direction has the form of

$$\Delta E = \int_{\xi_n}^{\xi_{n+1}} \frac{dH_z}{d\xi} d\xi = \int_{\xi_n}^{\xi_{n+1}} \frac{a_0^2 \sin 2\xi}{2\delta_B} d\xi. \tag{27}$$

As we can see, the integrand in Eq. (27) is smaller than that in the expression (20) no matter what form of the magnetic field is. On the other hand, $d\delta_B/d\xi > d\delta/d\xi$ such that $\delta_B$ grows faster out of the nonadiabatic region boundary, i.e., $\delta_B \sim 1$, which indicates that the contributing integral region in Eq. (27) is smaller than that in (20). As a result, we can conclude that the energy change in Eq. (27) is smaller than that in Eq. (20) with the appearance of the magnetic field and thus the stochastic heating is weakened.

For the case where laser polarized across to the magnetic field, the situation is more complex. This is because the energy variation also contains, apart from the integral in Eq. (27),

$$\Delta E = \int_{\xi_n}^{\xi_{n+1}} \frac{a_0 A_B \cos \xi}{2\delta_B} d\xi. \tag{28}$$

The large electron momentum $p_y = A_B$ along the laser electric field makes it possible to gain more energy from laser, which due to one collision is the combination of the integrals in Eq. (27) and (28). Assume that the effect due to the magnetic field is finite such that the electron still gets most acceleration in the vicinity of $\tilde{z}_{max}$ and $\delta_B \lesssim 1$ as seen in Sec. III. If $A_B(\tilde{z}_m)$ is small such that the contribution to electron energy variation from Eq. (28) is smaller than that from Eq. (27) and thus negligible, the stochasticity could be decreased like the case where the laser polarizes along the magnetic field. However, if the magnetic field is strong such that $A_B(\tilde{z}_m) > a_0$, the electron energy change is then determined by Eq. (28) and the stochastic motion could be enhanced. The study in Ref. 37 is an asymptotic limit of this case. The numerical simulations have been performed which agrees with these results.

## V. Results of numerical simulations

In order to confirm the analyses in the previous sections, we have numerically solved the 3/2D Hamiltonian equations for different laser and electric field parameters. As mentioned above, the upper limit of the stochastic energy in both transverse and longitudinal electric fields would increase for small electric field strength. This is not surprising since the weaker electric field allows the electron to stay in the nonadiabatic region for longer time such that it could gain more energy from the laser wave. However, from the point view of the stochastic heating, the requirements of the electric field strength are different in the transverse and longitudinal cases.

In the transverse case, a weak electric field should be employed to have $\varepsilon \ll 1$ since the stochastic heating occurs at high harmonic resonances $\Omega = \varepsilon\sqrt{E/E_{pond}} = \sqrt{E/E_{max}^{abs}} \ll 1$ and we are interested in electron energy accelerated above the ponderomotive scaling. For the electron with large energy such that $\Omega \tilde{<} 1$, the strong interaction of electron with laser pulse could take place along the whole electron orbit in Fig. 1 ($|\theta| \sim 1$) and the analytical results in Eq. (12-17) is not accurate and the laser impact is not like random "kicks". For example, Fig. 2 (which looks like Fig. 10 in Ref. 41) shows the electron dynamics for $C_\perp = 1$, $\kappa_u = 0.01$, y-polarized laser with amplitude $a_0 = 1$ and initial energy $E = 13$. For these parameters, $\Omega \sim 1$ for large electron energy $E \sim 100$ where the electron wanders near the boundary of stochasticity and its dynamics is complex. Therefore, in order to check the result in Eq. (17) which only holds for $\Omega \ll 1$ and thus only $|\theta| \ll 1$ contributes to the electron energy variation, we requires $\Omega_{max} = \sqrt{E_{max}^{x,y}/E_{max}^{abs}} \approx \varepsilon^{1/2} \ll 1$ and take $\kappa_u \leq 10^{-4}$, $a_0 \sim 1$ and $C_\perp = 1$ such that $\varepsilon \sim 0.01$. On the other hand, the electron oscillating frequency in Eq. (22) for the longitudinal electric field case inversely depends on the electron energy such that large electric field coefficient $k_u \tilde{<} \min\{a_0, a_0^{2-p}\} = \min\{a_0, 1\}$ could also cause strong stochastic electron heating. In the simulations we specify the static electric field as $U = k_u \tilde{z}^2 / 2$ and thus consider $k_u < \min\{a_0, a_0^{2-p}\} = \min\{a_0, 1\}$. Although smaller $k_u$ can lead to larger stochastic electron energy, $k_u \sim 0.1$ would be used in the simulations for time saving consideration which is sufficient to check the results in Sec. III.

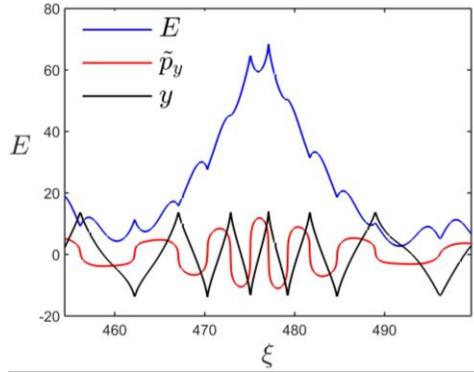

FIG. 2. Electron motion in the transverse electric field for $C_\perp = 1$, $\kappa_u = 0.01$, y-polarized laser amplitude $a_0 = 1$ and initial energy $E = 13$.

In the transverse case, the laser polarization is of great importance for electron dynamics since the electron motion is more stochastic for laser polarized along the quasi-static electric field than that for laser across to the electric field. Taking into account that $\Omega_{max}^x \approx \varepsilon^{6/11} < \Omega_{max}^y \approx \varepsilon^{3/7}$, the latter case is preferable to distinguish the stochastic heating. Therefore, we will present only the numerical results for the laser polarization of $\tilde{\vec{A}} = a_0 \sin(\xi)\vec{e}_x$ in the transverse case. All the simulations are set to $C_\perp = 1$. However, the laser polarization for electron in the longitudinal electric field is not important so we also take $\vec{\tilde{A}} = a_0 \sin(\xi)\vec{e}_x$. Moreover, all the analytic results are applied to the electron with $E > E_{pond}$ such that preheated relativistic electrons are placed at $y = 0$ and $\tilde{z} = 0$ in the simulations for the transverse and longitudinal cases, respectively. This is especially important for the transverse case due to the existence of the threshold-type dependence of the final energy gain[29] on the parameter of $\varepsilon$.

The Poincaré mappings in numerical simulations are formed in the following way: it's on 2D energy E and laser phase $\Delta\xi$ ($0 < \Delta\xi < \pi$) space where $\Delta\xi \equiv \xi_n - m\pi$ with $m \equiv [\xi_n / \pi]$. $E_n$ and $\xi_n$ are picked from the center of nonadiabatic regions. In Fig. 3 we have shown the numerical results of the maximum stochastic energy of electrons, in unit of $E_{max}^{abs}$, picking from the Poincaré mappings as a function of the parameter $\varepsilon$ in the logarithmic diagram as well as their fitting by linear polynomials for the transverse electric field case. The blue squares correspond to the data of $\kappa_u = 10^{-4}$ and varying $a_0$ while the red diamonds are for $a_0 = 8$ and varying $\kappa_u$. As we can see, the numerical simulations agrees with the analytic results in Eq. (17). The corresponding results for the longitudinal case are plotted in Fig. 4 to compare with (26).

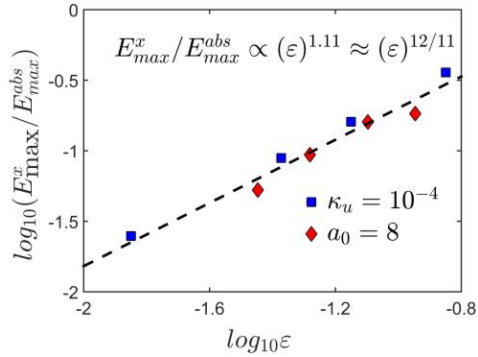

FIG. 3. The maximum stochastic energy $E_{max}^x / E_{max}^{abs}$ versus $\varepsilon$ of electrons in the transverse electric field for laser polarized across to the electric field in the logarithmic diagram and its fittings by a linear polynomial (the blue squares are for $\kappa_u = 10^{-4}$ and varying $a_0$ whereas the red diamonds are for $a_0 = 8$ and different $\kappa_u$).

The inaccuracy in these fittings reflects the difficulty to determine the maximum stochastic energy from the Poincaré mappings. This is because, when the stochastic parameter K closes to unity[37], "the structure of the phase space becomes complicated where the fraction of stable components of the motion plays an important role". Fig. 5 has displayed Poincaré mapping for electron in the transverse (in the top panel) and longitudinal (in the bottom panel) electric fields as examples. The large excursion at high energy is a well-known phenomenon at the boundary of stochasticity. In the simulations, the energy below the large excursion in the transverse case has been taken as the maximum stochastic energy.

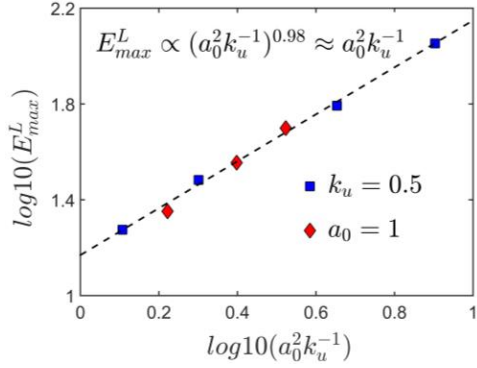

FIG. 4. The maximum stochastic energy $E_{max}^L$ versus $a_0^2 k_u^{-1}$ of electrons in the longitudinal electric field described by $U = k_u \tilde{z}^2 / 2$ in the logarithmic diagram and its fittings by a linear polynomial (the blue squares are for $k_u = 0.5$ and varying $a_0$ whereas the red diamonds are for $a_0 = 1$ and different $k_u$).

The Poincaré mapping for the longitudinal case in the bottom panel of Fig. 5 exhibits a series of stability islands not occupied by the electron trajectories. Their physics can be studied as following. The stationary points in the phase space are determined according to Eq. (24) by the solutions of

$$Q \sin \psi_0 = 0, \ \Pi_0^{1/3} = N\pi, \tag{29}$$

where N is integer. Therefore we have $\psi_0 = 0$ or $\psi_0 = \pi$ and $\Pi_0 = 8^3 k_u^{-3/2} E^{3/2} = (N\pi)^3$. The stability of these stationary points are determined by the eigenvalues of the Jacobian of the map[35,37] in the neighborhood of $\psi_0$ and $\Pi_0$

$$\begin{bmatrix} \partial \Pi_{n+1} / \partial \Pi_n & \partial \Pi_{n+1} / \partial \psi_n \\ \partial \psi_{n+1} / \partial \Pi_n & \partial \psi_{n+1} / \partial \psi_n \end{bmatrix} = \begin{bmatrix} 1 & Q \cos \psi_0 \\ \Pi_0^{-2/3} / 3 & 1 + \Pi_0^{-2/3} Q \cos \psi_0 / 3 \end{bmatrix}. \tag{30}$$

The stability condition requires that

$$|2 + 3^{-1} \Pi_0^{-2/3} Q \cos \psi_0| < 2. \tag{31}$$

Therefore the stationary points at $\psi_0 = 2n\pi$ (n is integer) is always unstable while for $\psi_0 = (2n+1)\pi$ the stability requires $\Pi_0^{-2/3} Q < 12$ which corresponds to

$$E > \Lambda a_0^2 / 2k_u. \tag{32}$$

It indicates that larger stability islands occurs at higher electron energy and $\psi_0 = (2n+1)\pi$ ($\Delta \xi = \pi/2$) as shown in the bottom panel of Fig. 5.

The same procedure can be applied to the transverse case, where for the laser polarized in x-direction, we have

$$|2 - 3/8 (\Pi_0^x)^{-11/8} Q_x \cos \psi_0^x| < 2, \ Q_x \sin \psi_0^x = 0, \text{ and } (\Pi_0^x)^{-3/8} = N\pi. \tag{33}$$

As a result, $\psi_0^x = (2n+1)\pi$ ($\Delta \xi = \pi/2$) are always unstable stationary points and the points of $\psi_0^x = 2n\pi$ ($\Delta \xi = 0$ or $\Delta \xi = \pi$) are stable for

$$E > E_{max}^x, \tag{34}$$

where a factor of order of unity has been omitted. Therefore, the stability island is present only for electron energy above the completely stochastic region as shown in the top panel of Fig. 5.

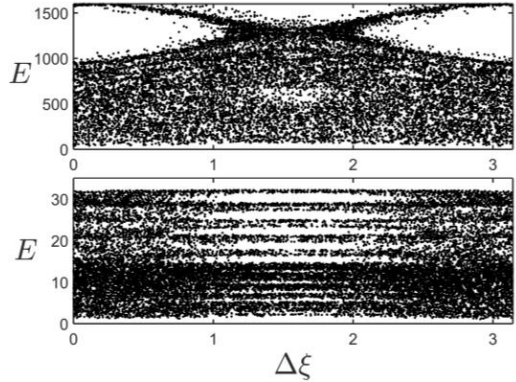

FIG. 5. Poincaré mappings for electrons in the transverse electric field in the top panel for $a_0 = 8$ and $\kappa_u = 5 \times 10^{-5}$ and in the longitudinal field in the bottom panel for $a_0 = 1$ and $k_u = 0.5$.

Fig. 6 has shown the electron trajectories and energy in the transverse (part (a)) and longitudinal (part (b)) electric fields, respectively. As one can see, it agrees with the analytic results in the previous sections, including the electron oscillating frequency, electron trajectories allowing long tail spectrum and its diffusive energy ("kicks") for highly stochastic motion, etc. For both cases, the variation of electron energy occurs primarily within a certain time interval much smaller than its characteristic periods. For the transverse case, the energy change mainly occurs around $y_{max}$ and $y_{min}$ where $|\theta| \ll 1$ corresponds to the energy jump. In the longitudinal case, the electron energy variation mainly comes from $\delta \ll 1$ and $\tilde{z} \approx \tilde{z}_{max}$, which is a factor of 2~3 larger than that $\delta \ll 1$ and $\tilde{z} \approx 0$. The electron motion in the transverse field as shown in Fig. 6(a) experiences relatively larger oscillations in the adiabatic region compared with that in the longitudinal case. This is because in the transverse case the dephasing rate $\gamma - p_z = C_\perp - U$ is always smaller than unity ($C_\perp$) such that even $|\theta| \sim 1$ region could slightly change the electron energy, while in the longitudinal case the dephasing rate $\gamma - p_z = \delta$ can be much larger than unity [see Fig. 6(b)] such that the energy variation is quite smaller.

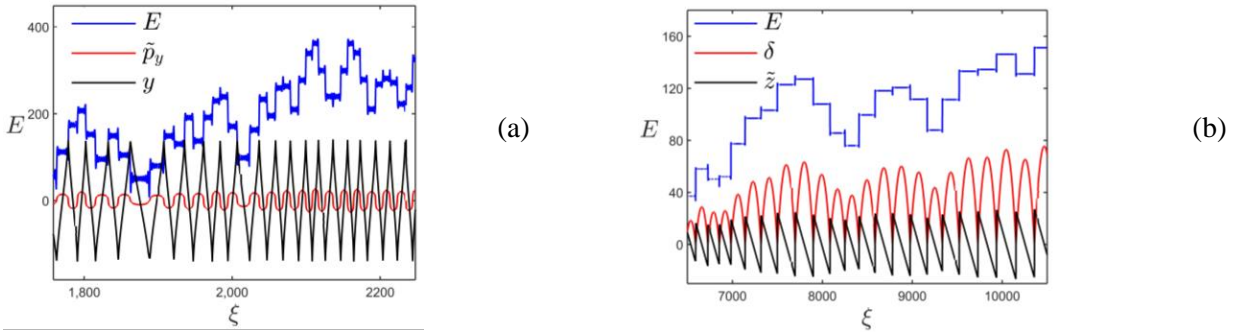

FIG. 6. Electron motions in the (a) transverse electric field for $C_\perp = 1$, $a_0 = 5$ and $\kappa_u = 10^{-4}$ and in the (b) longitudinal electric field for $a_0 = 5$ and $k_u = 0.1$. To make them readable, the canonical coordinates in both cases have been shrunk by some factors to illustrate their shapes.

To confirm the role of superluminal phase velocity, numerical simulations for the same parameters with Fig. 5 but $\alpha=1.001$ for electron in the transverse electric field and $\alpha=1.01$ in the longitudinal one are exhibited in Fig. 7. Compared with Fig. 5, it demonstrates that the superluminal phase velocity indeed weakens the stochastic heating in both cases but to different degrees. This seems to agree with the analyses of the roles of the superluminal phase velocity in the transverse and longitudinal cases, where it significantly reduces the dephasing rate for electrons in the transverse field via $\gamma - p_z = C_\perp - U(y) + (\alpha-1)p_z$ and thus the stochasticity, but only introduces some modest oscillations in Eq. (20) for the longitudinal case.

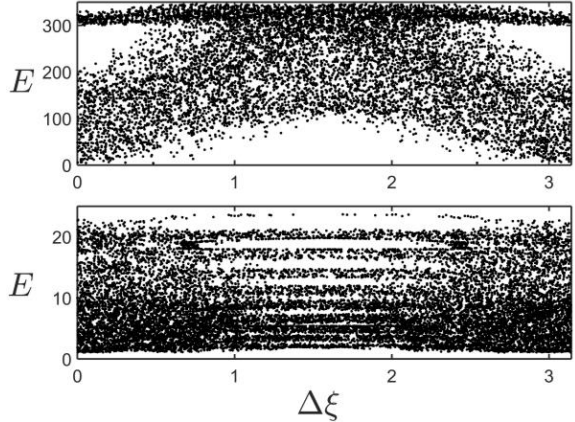

FIG. 7. Poincaré mappings for electrons for the same parameters with Fig. 5 but $\alpha=1.001$ for the transverse electric field in the top panel and $\alpha=1.01$ for the longitudinal case in the bottom panel.

## VI. Discussions and conclusions

The dynamics of ultrarelativistic electrons in the intense laser radiation and quasi-static electromagnetic fields has been studied in the 3/2 dimensional Hamiltonian framework. By using the stationary-phase method, Poincaré mappings are derived for electrons in both transverse and longitudinal electric fields, where the criterions for the onset of stochasticity are obtained. It is shown that for the linearly polarized laser radiation in the plane wave form, there exist upper limits of the stochastic heating in both cases. These maximum stochastic energies will be enhanced by larger laser intensity but weaker electric field.

We have found that the stochastic electron dynamics in the transverse electric field are quite different from that in the longitudinal one. In the first place, for the transverse case the unperturbed electron oscillation frequency is increasing with the increase of the electron energy such that lower harmonics resonances are achieved for electron near the boundary of stability. It was shown that the maximum stochastic electron energy depends only on the ponderomotive scaling and a small parameter of $\varepsilon = \sqrt{2} a_0 \kappa_u^{1/2} C_\perp^{-2/3}$. The situation in the longitudinal case is different where high harmonics resonances take place at large electron energy. Secondly, the role played by the electric field is different. The presence of the static electric potential in the transverse case directly decreases the dephasing rate such that the strong electron laser interaction occurs when electron is climbing the potential well. However, such strong interaction region in the longitudinal case locates at the bottom of the electrostatic potential well and thus the longitudinal electric field seems to provide only the confinement which forces the electron to enter the nonadiabatic region multiple times. Moreover, the polarization of the laser is of great importance in the transverse electric field but not in the longitudinal one. In the Poincaré mappings, the stability islands also behave differently.

The effects of the magnetic fields on these electron dynamics are discussed qualitatively. In the transverse case, $\kappa_u + \kappa_b$ plays the role of $\kappa_u$ such that the magnetic field can enhance, weaken, or terminate the stochastic electron motion depending on the sign and magnitude of $\kappa_b$. While in the longitudinal case, the stochasticity is always weakened by the magnetic field for the laser polarized along the magnetic field but can also be enhanced for proper magnetic field if the laser polarized across to it.

Numerical simulations to solve the 3/2D Hamiltonian equations directly have been performed, which have confirmed all the analytic results. The diffusive "kicks" of electron energy in the simulations indicate a highly stochastic motion of the electrons.

The stochastic heating requires high harmonic resonances, which requires $\Omega \ll 1$. As a result, the electron motion in the static fields takes a few or decades of laser periods to finish one oscillation. For example, to have the Poincaré mappings in Fig. 5 and 7 which contains $10^4$ points, we have $\xi \sim 10^5$. However, the electron could be heated to the order of maximum stochastic energy in a few oscillating periods depending on the electron initial conditions.

**Acknowledgments.** This work has been supported by the University of California Office of the President Lab Fee grant number LFR-17-449059.

**References**
[1] Tajima, T. and Dawson, J. M., Phys. Rev. Lett. **42,** 267(1979)
[2] Mendonça, J. T., Phys. Rev. A **28,** 3592(1983)
[3] Katsouleas, T. and Dawson, J. M., Phys. Rev. Lett. **51,** 392(1983)
[4] Rax, J.-M., Phys. Fluids B **4,** 3962(1992)
[5] Wilks, S. C., Kruer, W. L., Tabak, M., and Langdon, A. B., Phys. Rev. Lett. **69,** 1383(1992)
[6] McKinstrie C. J. and Startsev E. A., Phys. Rev. E **56,** 2130(1997)
[7] Pukhov, A. and Meyer-ter-Vehn, J, Phys. Plasmas **6,** 2847(1999)
[8] Gahn, C., Tsakiris, G.D., Pukhov, A., Meyer-ter-Vehn, J., Pretzler, G., Thirolf, P., Habs, D. and Witte, K.J., Phys. Rev. Lett. **83,** 4772(1999)
[9] Tanaka, K.A., Kodama, R., Fujita, H., Heya, M., Izumi, N., Kato, Y., Kitagawa, Y., Mima, K., Miyanaga, N., Norimatsu, T. and Pukhov, A., Phys. Plasmas **7,** 2014(2000)
[10] Tsakiris, G.D., Gahn, C. and Tripathi, V.K., Phys. Plasmas **7,** 3017(2000)
[11] Pukhov, A. and Meyer-ter-Vehn, J. Appl. Phys. B **74,** 355(2002)
[12] Shvets, G., Fisch, N.J. and Rax, J.M, Phys. Rev. E **65,** 046403(2002)
[13] Tanimoto, M., Kato, S., Miura, E., Saito, N., Koyama, K. and Koga, J.K., Phys. Rev. E **68,** 026401(2003)
[14] Esirkepov, T., Borghesi, M., Bulanov, S.V., Mourou, G. and Tajima, T., Phys. Rev. Lett. **92,** 175003(2004)

[15] Mangles, S.P., Walton, B.R., Tzoufras, M., Najmudin, Z., Clarke, R.J., Dangor, A.E., Evans, R.G., Fritzler, S., Gopal, A., Hernandez-Gomez, C. and Mori, W.B., Phys. Rev. Lett. **94,** 245001(2005)

[16] Lu, W., Huang, C., Zhou, M., Mori, W.B. and Katsouleas, T, Phys. Rev. Lett. **96,** 165002(2006)

[17] Mendonça, J.T., Silva, L.O. and Bingham, R., J. Plasma Phys. **73,** 627(2007)

[18] Geyko, V.I., Fraiman, G.M., Dodin, I.Y. and Fisch, N.J., Phys. Rev. E **80,** 036404(2009)

[19] Yabuuchi, T., Paradkar, B.S., Wei, M.S., King, J.A., Beg, F.N., Stephens, R.B., Nakanii, N., Hatakeyama, M., Habara, H., Mima, K. and Tanaka, K.A., Phys. Plasmas **17,** 060704(2010)

[20] Pollock, B.B., Clayton, C.E., Ralph, J.E., Albert, F., Davidson, A., Divol, L., Filip, C., Glenzer, S.H., Herpoldt, K., Lu, W. and Marsh, K.A., Phys. Rev. Lett. **107,** 045001(2011)

[21] Paradkar, B.S., Wei, M.S., Yabuuchi, T., Stephens, R.B., Haines, M.G., Krasheninnikov, S.I. and Beg, F.N., Phys. Rev. E **83,** 046401(2011)

[22] Mulser, P., Bauer, D. and Ruhl, H., Phys. Rev. Lett. **101,** 225002(2008)

[23] Arefiev, A.V., Breizman, B.N., Schollmeier, M. and Khudik, V.N., Phys. Rev. Lett. **108**, 145004(2012)

[24] Arefiev, A.V., Khudik, V.N. and Schollmeier, M., Phys. Plasmas **21,** 033104(2014)

[25] Robinson, A.P.L., Arefiev, A.V. and Neely, D., Phys. Rev. Lett. **111,** 065002(2013)

[26] Paradkar, B.S., Krasheninnikov, S.I. and Beg, F.N., Phys. Plasmas **19,** 060703(2012)

[27] Krasheninnikov S. I., Phys. Plasmas **21** 104510(2014)

[28] Arefiev, A.V., Robinson, A.P.L. and Khudik, V.N., J. Plasma Phys. **81,** 475810404(2015)

[29] Khudik, V., Arefiev, A., Zhang, X. and Shvets, G., Phys. Plasmas **23,** 103108(2016)

[30] Luan, S.X., Yu, W., Li, F.Y., Wu, D., Sheng, Z.M., Yu, M.Y. and Zhang, J. Phys. Rev. E **94,** 053207(2016)

[31] Wu, D., Krasheninnikov, S.I., Luan, S.X. and Yu, W., Nucl. Fusion **57,** 016007(2016)

[32] Wu, D., Luan, S.X., Wang, J.W., Yu, W., Gong, J.X., Cao, L.H., Zheng, C.Y. and He, X.T., Plasma Phys. Contr. Fusion **59,** 065004(2017)

[33] Peebles, J., Wei, M.S., Arefiev, A.V., McGuffey, C., Stephens, R.B., Theobald, W., Haberberger, D., Jarrott, L.C., Link, A., Chen, H. and McLean, H.S., New. J. Phys. **19,** 023008(2017)

[34] Chirikov B. V., Phys. Reports **52,** 263(1979)

[35] R. Z. Sagdeev, D. A. Usikov, and G. M. Zaslavsky, *Nonlinear Physics: From the Pendulum to Turbulence and Chaos* (Hardwood Academic Publishers GmbH, Switzerland, 1988).

[36] A. J. Lichtenberg and M. A. Lieberman, *Regular and chaotic Dynamics* (2nd Edition, Applied Mathematical Sciences, Vol. 38, New York, NY: Springer-Verlag, 1992).

[37] Zaslavskii, G.M., Natenzon, M.Y., Petrovichev, B.A., Sagdeev, R.Z. and Chernikov, A.A., Sov. Phys. JETP **66,** 496(1987)

[38] V. Khudik, Xi Zhang, Tianhong Wang, and Gennady Shvets, Phys. Plasmas **25**, 083101 (2018)

[39] Angus J. and Krasheninnikov S. I., Phys. Plasmas **16,** 113103(2009)

[40] Zhang Y. and Krasheninnikov S. I., Phys. Plasmas **25,** 013120(2018)

[41] Zhang Y. and Krasheninnikov S. I., Physics Letters A **382.27,** 1801(2018)

[42] Arefiev, A.V., Khudik, V.N., Robinson, A.P.L., Shvets, G., Willingale, L. and Schollmeier, M., Phys. Plasmas **23,** 056704(2016)